%% file: manuscript.tex
\documentclass[sigconf,screen,authorversion]{acmart}

\graphicspath{ {./image/} }

\usepackage{enumitem}
\setlist[itemize]{leftmargin=*}
\usepackage{graphicx}
\usepackage{caption}
\usepackage{subcaption}
\usepackage{placeins}
\usepackage{float}
\usepackage{multirow}
\usepackage{balance}
\usepackage{xspace}
\usepackage{adjustbox}

\copyrightyear{2023}
\acmYear{2023}
\setcopyright{acmlicensed}
\acmConference[SIGIR '23] {Proceedings of the 46th International ACM SIGIR Conference on Research and Development in Information Retrieval}{July 23--27, 2023}{Taipei, Taiwan.}
\acmBooktitle{Proceedings of the 46th International ACM SIGIR Conference on Research and Development in Information Retrieval (SIGIR '23), July 23--27, 2023, Taipei, Taiwan}
\acmPrice{15.00} 
\acmDOI{10.1145/3539618.3591981} 
\acmISBN{978-1-4503-9408-6/23/07}

\input{macros.tex}

\begin{document}

\title{Examining the Impact of Uncontrolled Variables on Physiological Signals in User Studies for Information Processing Activities}

\author{Kaixin Ji}
\orcid{0000-0002-4679-4526}
\affiliation{
\institution{RMIT University}
  \city{Melbourne}
  \country{Australia}
}
\email{kaixin.ji@student.rmit.edu.au}

\author{Damiano Spina}
\orcid{0000-0001-9913-433X}
\affiliation{
\institution{RMIT University}
  \city{Melbourne}
  \country{Australia}
}

\email{damiano.spina@rmit.edu.au}

\author{Danula Hettiachchi}
\orcid{0000-0003-3875-5727}
\affiliation{
\institution{RMIT University}
  \city{Melbourne}
  \country{Australia}
}

\email{danula.hettiachchi@rmit.edu.au}

\author{Flora Dilys Salim}
\orcid{0000-0002-1237-1664}
\affiliation{%
  \institution{The University of New South Wales} 
  \city{Sydney}
  \country{Australia}
}
\email{flora.salim@unsw.edu.au}

\author{Falk Scholer}
\orcid{0000-0001-9094-0810}
\affiliation{
\institution{RMIT University}
  \city{Melbourne}
  \country{Australia}
}
\email{falk.scholer@rmit.edu.au}

\setlength{\intextsep}{10pt plus 2pt minus 2pt}
\renewcommand{\shortauthors}{Kaixin Ji, Damiano Spina, Danula Hettiachchi, Flora Dilys Salim, \& Falk Scholer}

\begin{abstract}
Physiological signals can potentially be applied as objective measures to understand the behavior and engagement of users interacting with information access systems. However, the signals 
are highly sensitive, and many controls are required in laboratory user studies. To investigate the extent to which controlled or uncontrolled (i.e., confounding) variables such as task sequence or duration influence the observed signals, 
we conducted a pilot study where each participant completed four types of information-processing activities (\reading, \listening, \speaking, and \writing). Meanwhile, we collected data on blood volume pulse, electrodermal activity, and pupil responses. We then used machine learning approaches as a mechanism to examine the influence of controlled and uncontrolled variables that commonly arise in user studies. Task duration was found to have a substantial effect on the model performance, suggesting it represents individual differences rather than giving insight into the target variables. This work contributes to our understanding of such variables in using physiological signals in information retrieval user studies.

\end{abstract}

\settopmatter{printfolios=true}


\maketitle

\section{Introduction}

Laboratory user studies are an effective mechanism to understand how users interact with information access and retrieval systems~\cite{kelly2009methods, hearst2009search}. However, designing user studies is not trivial, as it requires controlling for numerous factors influenced by each individual participating in the experiment, and their perception of the information and task.
Information-seeking models have been proposed to characterize how users interact with a variety of systems such as screen-based search~\cite{belkin1980anomalous,marchionini1997information}, spoken conversational search~\cite{trippas2020towards}, multimedia platforms~\cite{white2016interactions}, or multi-modality input/output in conversational recommender systems~\cite{2022surveycrs}. Most user-system interactions in such systems involve four basic Information-Processing Activities (IPAs): \reading, \writing, \listening, and \speaking.
With the recent advances in wearable devices, it is natural to wonder what physiological signals can tell us about how users engage in these IPAs.
In this paper, we present the results of a laboratory user study ($N=7$) where physiological signals -- Electrodermal Activity (EDA), Blood Volume Pulse (BVP), and Pupil Diameter (PD) -- are collected using a wearable device and an eye-tracker. 
Our ultimate aim is to analyze the signals by defining a multi-class classification problem: Can we predict the specific IPA the user performed by feeding a machine-learning model with the signals obtained from the sensors? The signals are sensitive to noise. Therefore, we analyze how a set of variables (both controlled and uncontrolled, but likely to interfere) influence the machine-learning model's performance, and whether this can be used as a mechanism to scrutinize the validity of our experimental design. 
Data and code are publicly available online.\footnote{GitHub: \url{bit.ly/ji2023examining}}

The contributions of this paper are two-fold:
\begin{itemize}
    \item We introduce a simple but informative methodology that, by observing changes in the effectiveness of machine learning models, can potentially characterize the influence of controlled and uncontrolled variables in complex laboratory user studies.
    \item The results of this analysis revealed that some variables, such as the duration of the task, should be carefully designed and gauged before running the study at a larger scale.
\end{itemize}


\section{Related Work}
\label{sec:related}

User studies are widely used to understand how users interact with information access and retrieval systems and to collect data for evaluating such systems. `In-the-wild' studies, e.g., \cite{mcduff2021affect}, involved complicated factors. Thus lab studies should be cautiously conducted beforehand to understand target outcomes and conditions. On the other hand, there is growing interest in bio-signals (e.g., fMRI \cite{moshfeghi2016understanding}, eye-tracking \cite{granka2004eye, arapakis2009using, buscher2012attentive, cole2014task}), especially when looking at affective feedback \cite{wu2017predicting, White2017881}. Despite the sensitivity of bio-signals \cite{mostafa2016deepening}, the experiment can be influenced by various factors,  including task activities, task designs, environments, and participants' inner status (e.g., physical or mental state).  
\citet{moshfeghi2016understanding}’s user study consisted of two interaction activities; pressing a physical button to respond and verbally expressing a search query. The participants in \citet{arapakis2009using}’s study performed both video- and text-search tasks. \citet{lin2007multilayer} included listening tasks. 
\citet{granka2004eye} controlled the task difficulty levels and topics. \citet{moshfeghi2013effective} controlled the task types (search intentions). \citet{buscher2012attentive} controlled the document relevance in two user studies. The relevance varied across short documents in the first and across sections within a long document in the second. The fatigue and complexity might have confounding effects.

In affective computing, multi-modal data have been used to detect emotions with three or more classes. 
Both \citet{verma2014multimodal} and \citet{ganapathy2020convolutional} proposed architectures of deep learning feature-extraction methods with machine learning approaches, including Multilayer Perceptron (MLP), Support Vector Machines (SVM), and K-Nearest Neighbors (KNN) multi-label classifiers; while the former used multi-modal signals and the other used only EDA. 
They both tested with the DEAP database \cite{koelstra2011deap}. It contains multiple physiological signals and emotion annotations from watching short videos. A short break is provided after around 27 minutes. 
Another popular multi-modal emotion dataset is the CEAP-360VR dataset \cite{xue2021ceap} which contains physiological and behavioral signals.

In summary, it remains unclear if multiple activities or variables contained in one user study would impact the experiment results and require careful control in information processing experiments.
In this regard, we select the variables that are commonly encountered and report their impact on a rigorous laboratory study to infer whether they require careful treatment in the experimental design.

\section{User Study}
\label{sec:data}
\paragraph{Participants}
As a pilot study, 8 participants were recruited in total. However, data for one participant had to be discarded due to the collection error and data from 7 participants (5M, 2F) were analyzed.
One participant was in the 35--44 age group, and others were in the 25--34 age group. 
The study received ethics approval from the author's university, and the participants provided written consent before the experiment. 

\begin{figure}[htbp!]
 \setlength{\abovecaptionskip}{1pt}
     \centering
     \includegraphics[width=0.8\linewidth, trim={1cm 4cm 1cm 4cm}, clip=true]{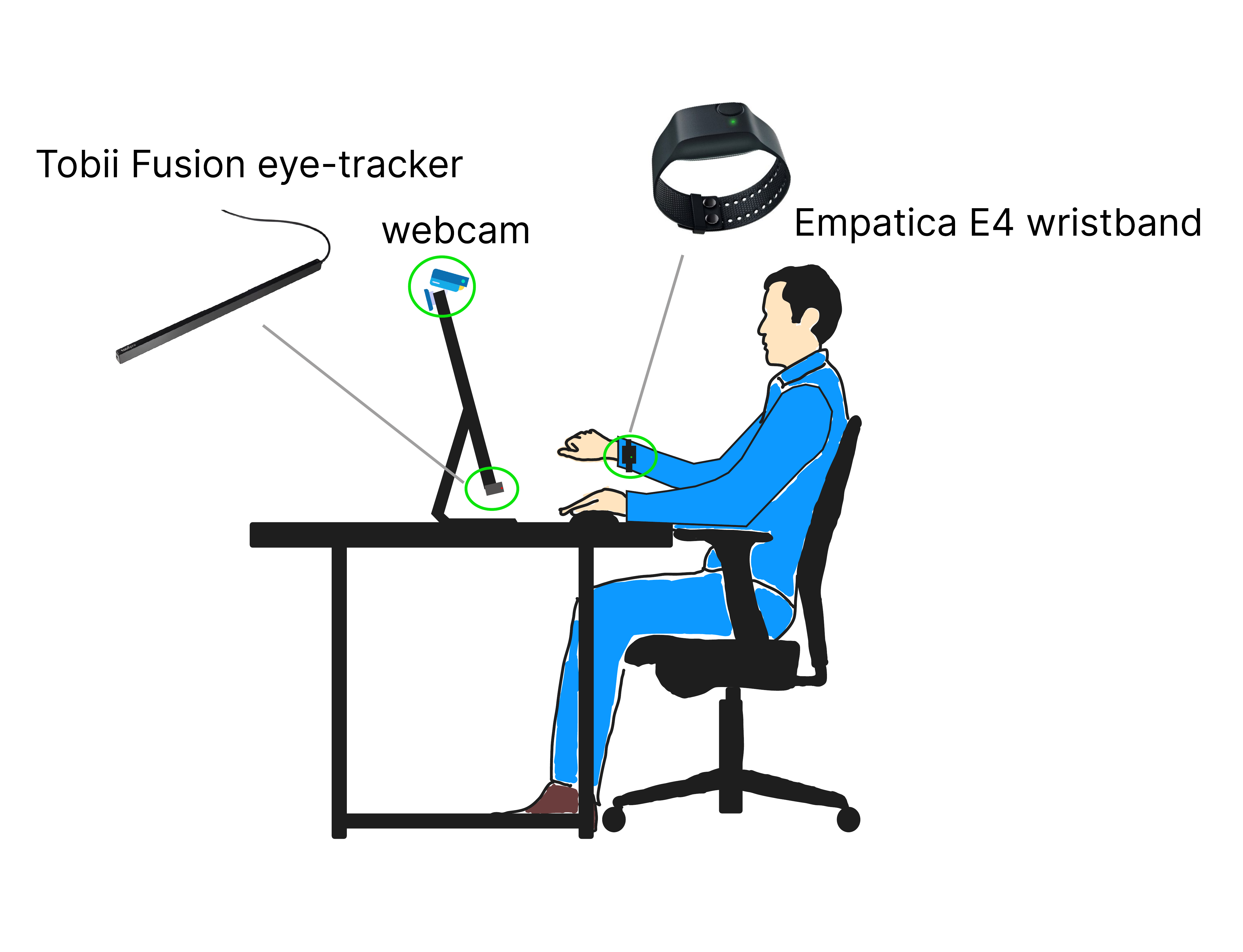}
     \caption{Experiment equipment setup.}
     \label{fig:set_up}
 \end{figure}

\paragraph{Setup and Equipment}
There are three sensors used in this study: a webcam camera for video recording, a Tobii Fusion eye-tracker\footnote{\url{https://www.tobii.com/products/eye-trackers/screen-based/tobii-pro-fusion}} for PD and an E4 wristband\footnote{\url{https://www.empatica.com/en-int/research/e4/}} for EDA and BVP. 
The equipment setup is shown in Figure~\ref{fig:set_up}.
There is a desktop PC mounted with an eye-tracker and a web camera in the experiment room. The participant sits in front of the computer and wears the wristband on the non-dominant hand. 
All participants used the computer mouse with their right hand. 
The instructor leaves the experiment room after calibration and instruction to avoid interruption. 

\raggedbottom

\begin{figure}[htbp!]
    \setlength{\abovecaptionskip}{1pt}
    \centering
    \includegraphics[width=\linewidth]{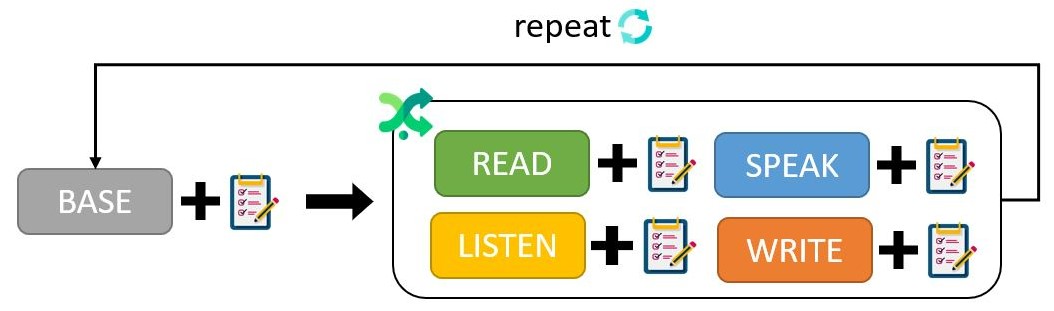}
    \caption{Flowchart of the user study. The sequences of tasks and complexity sections are randomized.}
    \label{fig:flowchart}
\end{figure}

\paragraph{Procedure}
Each participant first completes a pre-task survey. The survey asks for sleeping hours prior and caffeine intake on the experiment day, which might affect the cognition status~\cite{2021edachi}. 
Figure~\ref{fig:flowchart} presents the user study procedure. The study consists of two sections, each corresponding to two pre-defined activity complexity levels (low and high). For \reading and \listening IPAs, complexity is defined using low and high readability scores \cite{williams1972table}. The materials are scientific new items with around 500 words, some converted into synthesis speeches using Google Neutral Voice for \listening. For \speaking and \writing, complexity is estimated by the type of questions and the length of expected answers. The easy questions are recalling questions, e.g., `what was your routine this morning? (100 words minimum)'; the hard questions need recalling and analyzing, e.g., `does social media make you in general happier or sadder? why? (300 words minimum)'. Each section starts with a relaxing activity (\relaxing), where the participant is asked to watch a relaxing video and minimal cognitive efforts are involved. Then, the participant completes four IPAs: \reading, \listening, \speaking, and \writing. Specifically, the participant needs to read one article, listen to one article, answer a question by speaking, and answer a question by typing. After each task (including \relaxing), the participant completes an engagement scale~\cite{OBRIEN201828}. 

\paragraph{Data Pre-processing and Feature Extraction}
To synchronize all the sensors, we convert the timestamps into ISO 8601 time format with milliseconds. The signals are segmented according to the event timestamps recorded during the experiment. We follow similar data cleaning procedures as in~\cite{Lascio2018student,Bota2019,braithwaite2013guide} for EDA and BVP, and similar procedures as in~\cite{Zhai2005, kret2019preprocessing} for PD data. The pre-processed signals are first divided by sliding windows (2 seconds with 1-second overlap), then the average values of each feature are computed. The pre-processing steps and extracted features are described in Table~\ref{tab:features}.

\begin{table}[htbp!]
\resizebox{\linewidth}{!}{
\begin{tabular}{p{2.7cm}p{4.8cm}p{1.4cm}}
\toprule
\textbf{Data (Raw Hz)}                        & \multicolumn{1}{c}{\textbf{Pre-processing}}                  & \textbf{Features}                                   \\ 
\midrule
Electrodermal  & 1. Rolling median (5 sec. window)   & original,\\
Activity (EDA, 4 Hz)                                       & 2. Butterworth bandpass filter (1--8 Hz, $4^{\text{th}}$ order) & $1^{\text{st}}$ \& $2^{\text{nd}}$ derivative                                                         \\ 
Blood Volume Pulse                                       & 3. Min-max Normalization                             &                                                          \\
(BVP, 64 Hz)                                       & 4. EDA upsampled to 8 Hz                              &                                                          \\ 
\midrule
Pupil Diameter            & 1. Remove with range (1.5--9 mm)   & mean,       \\
(PD, 250 Hz)                                       & 2. Remove and correct the gap caused by blink      &     median,                                                 \\
                                       & 3. Linear Interpolation                              &           standard                                                     \\
                                       & 4. Combine two sides by averaging            &  deviation                                                        \\
                                       & 5. Zero-phase lowpass filter (4 Hz)                &                                                          \\
                                       & 6. Downsampled to 100 Hz                              &                                                          \\ 
\bottomrule
\end{tabular}}
    \caption{The pre-processing steps and extracted features for each signal. The sliding window is 2 seconds with 1-second overlap. Features are extracted as in \cite{xue2021ceap,feng2018wavelet}.}
    \label{tab:features}
\vspace{-5mm}
\end{table}
\raggedbottom

\section{Results And Analysis}
\label{sec:experiments}

Our classification task consists of classifying four activities: \reading, \listening, \speaking, and \writing.
Overall, our experiment follows a similar setting to the one proposed by~\citet{xue2021ceap}: it includes the same short-duration signals (70 seconds), the same sensor data (EDA, BVP, PD), the same collection devices for EDA and BVP data, and it addresses a multi-class classification problem. We use a leave-one-\emph{participant}-out approach for cross-validation. As we have data from 7 participants, we have 7 times cross-validation. Note that this is a robust way to split the data: all the folds are equally balanced (12 training and 2 test instances for each IPA), and on each fold, all the IPAs of the test participant are in the test split, minimizing the risk of learning individual patterns from participants.

\subsection{Model Selection}
\begin{figure}[htbp!]
\setlength{\abovecaptionskip}{1pt}
    \centering
    \includegraphics[width=0.8\linewidth]{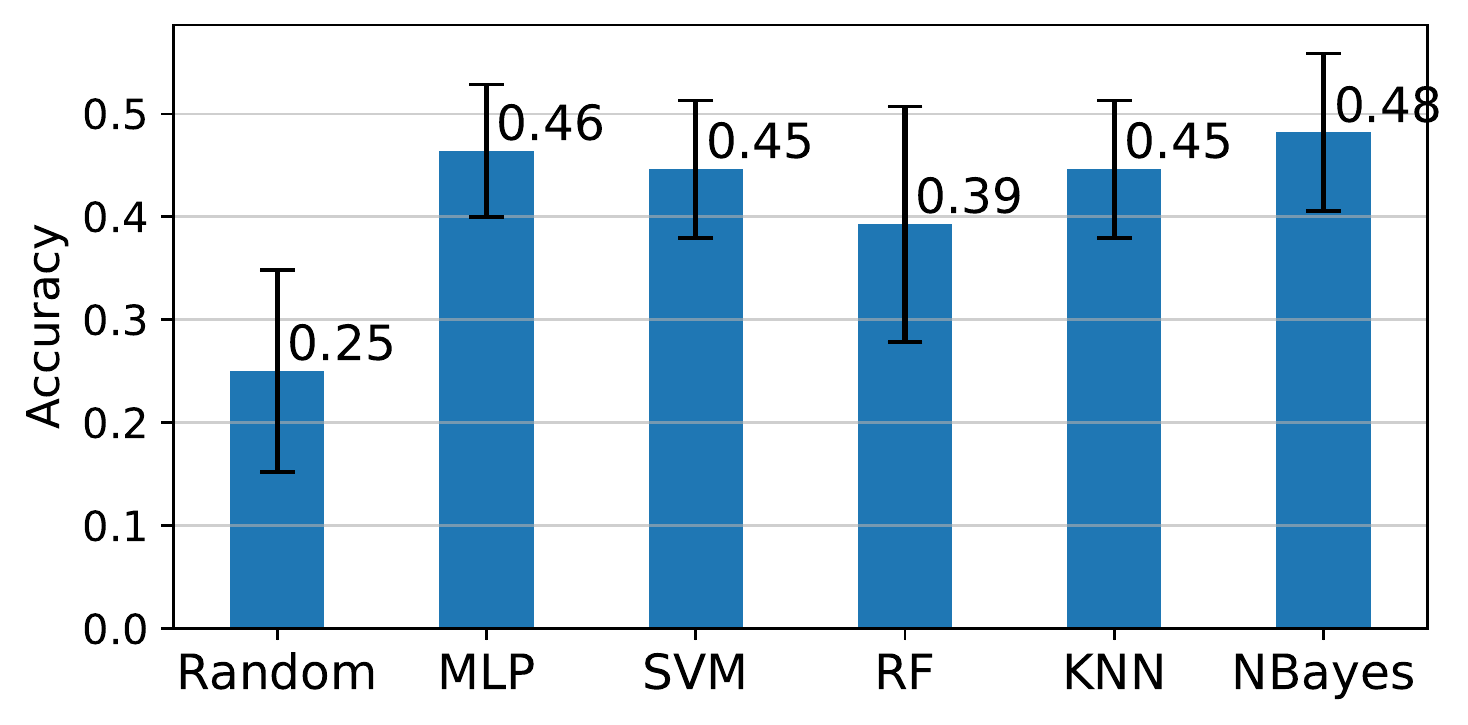}
    \caption{Accuracy of different multi-class machine learning models, using leave-one-participant-out (7 folds). Error bars indicate 95\% confidence intervals (\textit{t}-distribution).}
    \label{fig:model_compare}
\end{figure}

First, we compare different machine learning (ML) models which are commonly used on physiological data \cite{verma2014multimodal,ganapathy2020convolutional,xue2021ceap,MAITHRI2022106646,lin2010eeg,Gao2020}. 
Specifically, we use the following ML models: non-linear SVM (radial basis function kernel), RF (max. depth=4), KNN (Euclidean distance) and Na\"ive Bayes (NBayes). 
We also use MLP with the rectified linear unit (ReLU) activation function and Adam optimizer \cite{babu2018multimodal}. We also report the effectiveness of a Random classifier. Figure~\ref{fig:model_compare} shows the accuracy of different ML models for our 4-class IPAs classification.

\raggedbottom

Overall, we can see that ML models are able to learn from physiological signals with comparable effectiveness. NBayes obtains the highest mean accuracy, whereas KNN and SVM obtain a comparable performance with smaller confidence intervals (CI). 
We report the rest of the experiment using SVM, a non-probabilistic, effective model often used for physiological data classification \cite{xue2021ceap, RAJOUB202051, bakkialakshmi2021survey}. Similar trends were observed with the other ML methods we used.

\subsection{Impact of Variables}
In total, we examine the influence of five variables. They are two controlled variables -- \textit{`complexity'} and \textit{`task\_sequence'} -- and three uncontrolled variables -- \textit{`duration'}, \textit{`engage\_score'}, and
\textit{`cumulative\_time\_spent'}. The hypotheses are made based on the influence of each variable w.r.t. model effectiveness, which then can be used to inform changes to make the experimental design of our user study more robust. Intuitively, the model is supposed to calibrate the inputs (i.e., performance should not change) if the variables do not impact the user study.
We report confusion matrices that aggregate the results from the 7 leave-one-participant-out folds. Our test data also has balanced classes (8 IPAs per participant).
The prediction results with only signal features are presented in Figure~\ref{fig:signal_cm}. The model skews to \listening and \writing while less on \speaking.  

\begin{figure}[htbp!]
\setlength{\abovecaptionskip}{1pt}
    \centering
    \includegraphics[width=0.8\linewidth]{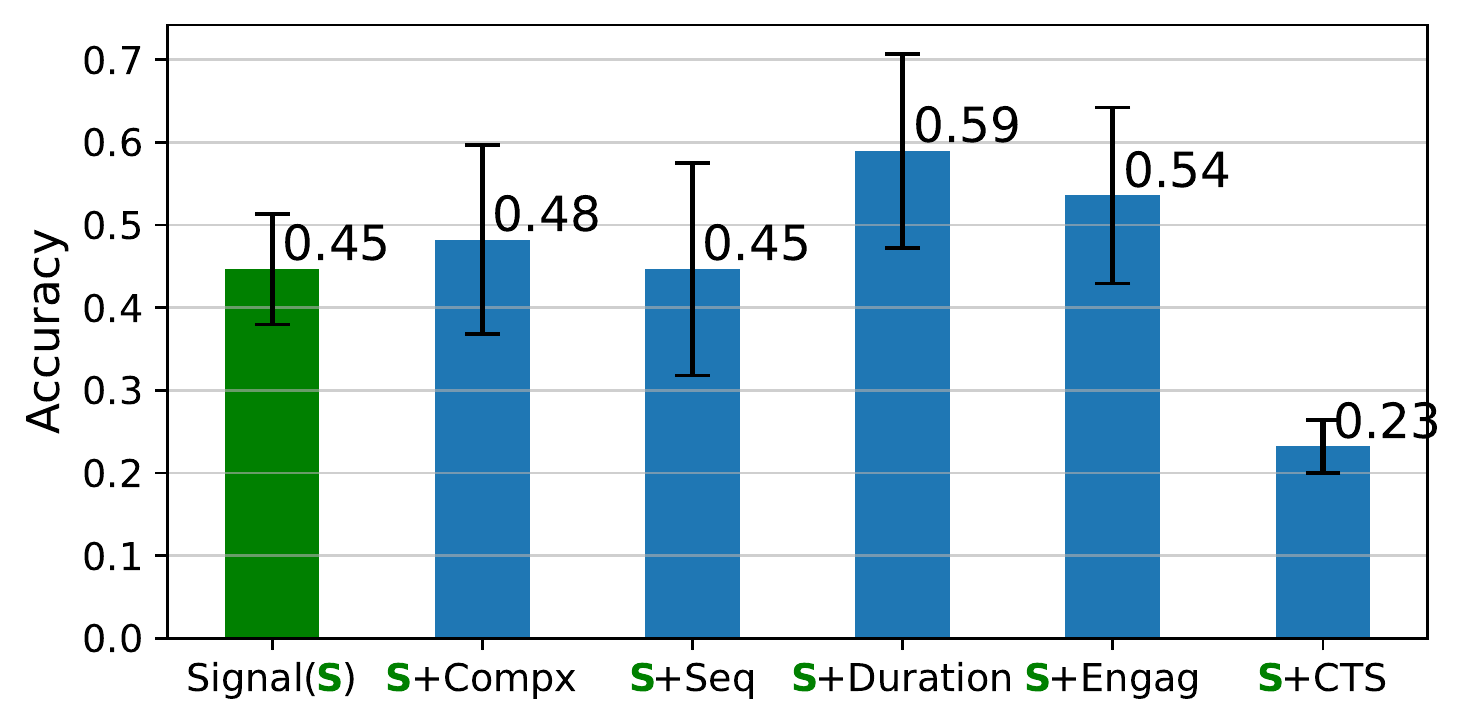}
    \caption{Accuracy for non-linear SVM classifiers, trained using signal features with one variable feature. CTS is the `cumulative\_time\_spent'. Error bars indicate 95\% confidence intervals (\textit{t}-distribution).}
    \label{fig:variable_bar}
\end{figure}

\subsubsection{Complexity}
\label{sec:complexity}
\begin{figure}[ht!]
\setlength{\abovecaptionskip}{1pt}
    \centering
    \begin{subfigure}[b]{0.34\linewidth}
    \setlength{\abovecaptionskip}{1pt}
        \includegraphics[width=\linewidth, height=26mm]{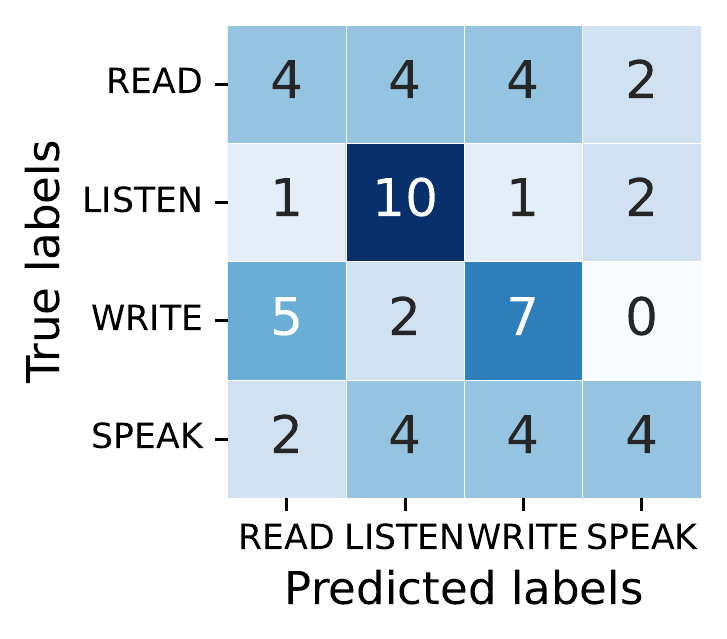}
        \caption{Signals}
        \label{fig:signal_cm}
    \end{subfigure}
    \begin{subfigure}[b]{0.32\linewidth}
    \setlength{\abovecaptionskip}{1pt}
        \includegraphics[width=\linewidth, trim=8mm 0 0 0, clip, height=26mm]{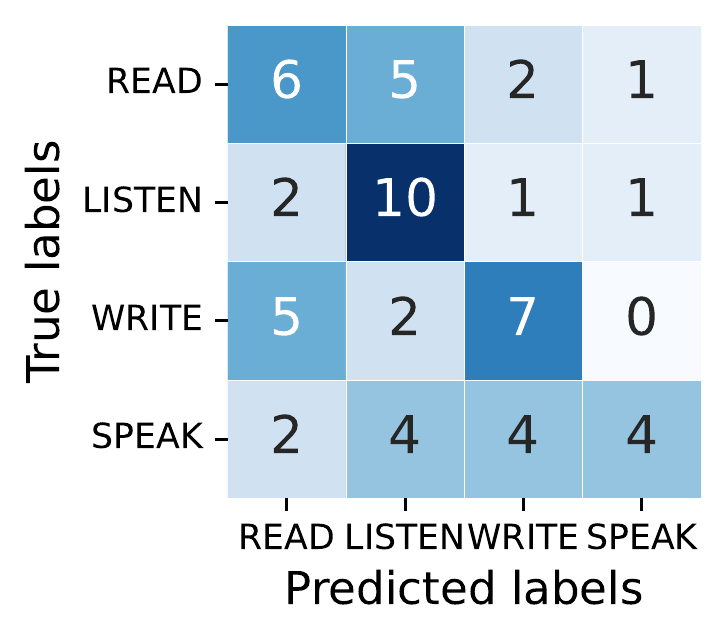}
        \caption{Complexity}
        \label{fig:comlexity_cm}
    \end{subfigure}
    \begin{subfigure}[b]{0.31\linewidth}
    \setlength{\abovecaptionskip}{1pt}
        \includegraphics[width=\linewidth, trim=8mm 0 0 0, clip, height=26mm]{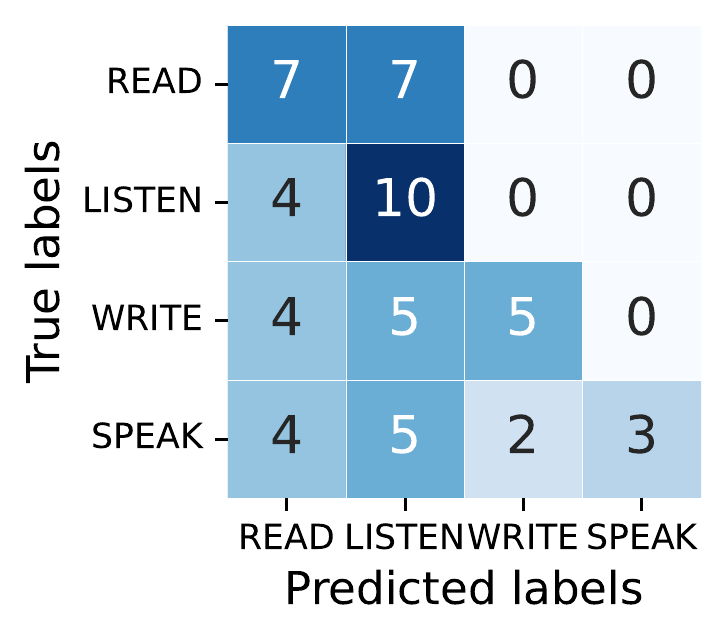}
        \caption{Task\_sequence}
        \label{fig:sequence_cm}
    \end{subfigure}
    \caption{Aggregated Confusion Matrix for the (a) Physiological Signals and controlled variables: (b) Complexity; and (c) Task Sequence.}
    \label{fig:confusion_matrix_control}
\end{figure}

In the user study, there are two sections for low or high complexity. We control this variable to prevent participant distraction from task difficulty.
As presented in Figure~\ref{fig:variable_bar}, the accuracy increases slightly after adding \textit{`complexity'} as a feature into the model. However, in addition to the confusion matrix in Figure~\ref{fig:comlexity_cm}, the complexity does not impact the classifications on \writing and \speaking at all, while it impacts classifications between \reading and \listening.
Furthermore, we conduct another experiment to classify the LOW and HIGH complexity using the signal features and linear SVM. The model results in low performance, 44.6\% accuracy ($\pm 10.2\%$), 36.9\% F1 ($\pm 11.3\%$), and 40.2\% AUC ($\pm 22.6\%$), indicates that the model is not able to classify complexities.

The results suggest the complexity level should be carefully controlled when designing a user study related to reading and listening. 
A limitation here is that complexity is instantiated in different ways for different IPAs (readability score for \reading and \listening, complexity of questions and expected length of answers for \speaking and \writing). The result of complexity impacts \reading and \listening but not \speaking and \writing, potentially due to this difference.

\subsubsection{Task Sequence}
\label{sec:sequence}
The \textit{`task\_sequence'} in our user study is randomized to counterbalance any potential order effect \cite{kelly2009methods}.
After adding the \textit{`task\_sequence'} as a feature, the accuracy does not change while the CI expands (Figure~\ref{fig:variable_bar}). In addition to Figure~\ref{fig:sequence_cm}, the model skews to the \reading or \listening more (82\%) compared to using only signals. However, the reason is probably the small number of participants; the sequence of 4 tasks (in 2 sections) cannot be counterbalanced. By design, \relaxing activities are always taken as first and sixth in the sequence. In our pilot study, only \speaking has equal occurrences, while \reading occurs mostly as last, \listening never appears as last, and \writing occurs in five out of the eight possible positions. 

The results suggest that \textit{`task\_sequence'} may not impact the model, as we have not observed changes in the model effectiveness when adding task\_sequence as a feature.
However, given the limited coverage of the possible sequence permutations in the existing data, we may observe different results in a larger dataset.


\begin{figure}[ht!]
\setlength{\abovecaptionskip}{1pt}
    \centering
    \begin{subfigure}[b]{0.34\linewidth}
        \setlength{\abovecaptionskip}{1pt}
        \includegraphics[width=\linewidth, height=26mm]{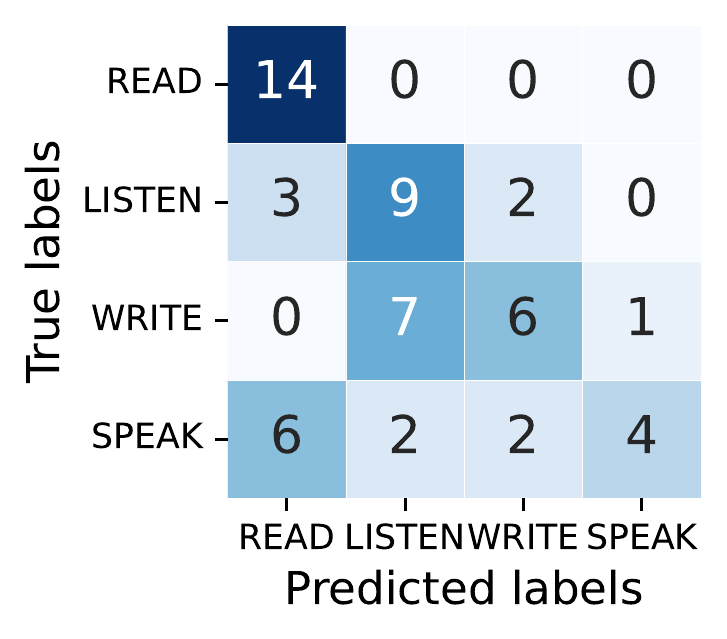}
        \caption{Duration}
        \label{fig:duration_cm}
    \end{subfigure}
    \begin{subfigure}[b]{0.33\linewidth}
        \setlength{\abovecaptionskip}{1pt}
        \includegraphics[width=\linewidth, trim=8mm 0 0 0, clip, height=26mm]{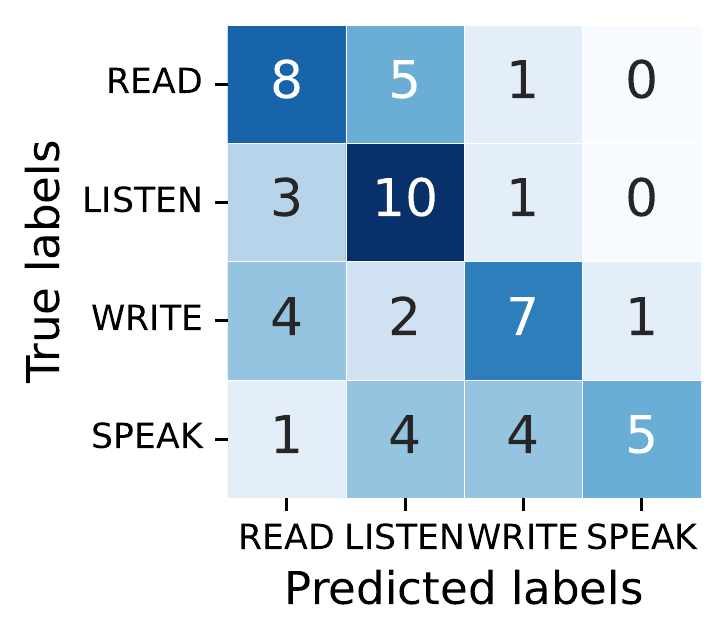}
        \caption{Engagement}
        \label{fig:engage_cm}
    \end{subfigure}
    \begin{subfigure}[b]{0.31\linewidth}
        \setlength{\abovecaptionskip}{1pt}
        \includegraphics[width=\linewidth, trim=8mm 0 0 0, clip, height=26mm]{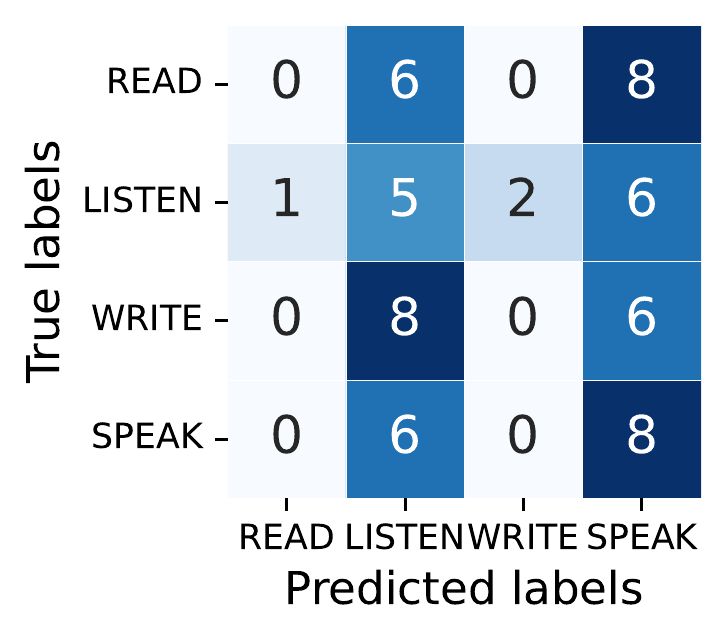}
        \caption{CTS}
        \label{fig:fatigue_cm}
    \end{subfigure}
    \caption{Aggregated Confusion Matrix for uncontrolled variables: (a) Duration; (b) Engagement; and (c) Cumulative Time Spent (CTS).}
    \label{fig:confusion_matrix_confound}
\end{figure}

\subsubsection{Activity Duration}
\label{sec:duration}
In our collected data, the duration unexpectedly varies. With this not controlled, we expect that \textit{`duration'} does not impact the effectiveness of the ML model.
However, the accuracy increases 10\% after adding the \textit{`duration'} as a feature. Both \reading and \writing are not time-restricted; the duration depends on the individual. But according to Figure~\ref{fig:duration_cm}, \reading has a 100\% True-Positive rate which indicates that \reading has a similar duration across participants, while \writing has more False-Negative results. In contrast, \speaking has more False-Negative results which are unanticipated. The `next' button only appears after a minimum pre-defined time, thus it should have fewer variants across participants, and so does the \listening, which the participants require to listen to audios for around 2:30 to 3 minutes. 

The results suggest that IPA \textit{`duration'} may play a strong role in the user study. This finding informs the refinement of our design: we will encourage duration consistency across tasks by informing users about the expected duration of each IPA, and by showing a timer (but not enforcing termination) to nudge users to complete the activity in time.

\subsubsection{Engagement}
\label{sec:engage}
During our user study, the participants report their engagement score \cite{OBRIEN201828} after completing each IPA. We round the engagement score into integers, thus it is a rank number from 1--5. Intuitively, different engagement levels would be reflected in the physiological signal. However, this should be independent of the actual type of IPA performed by the participants. 
After adding the \textit{`engagement'} as a feature, the model increases around 9\% accuracy but with a larger CI. As shown in Figure~\ref{fig:engage_cm}, engagement slightly increases the True-Positive for \reading and \speaking and the False-positive for \reading and \listening.

\subsubsection{Cumulative Time Spent}
\label{sec:fatigue}
In our user study, the whole session is expected to take 1 hour. We minimize the effect of fatigue by starting each section with a relaxing activity and providing a break time in the middle. 
\textit{`cumulative\_time\_spent'} is the cumulative time each participant takes till each IPA. It can be seen as a proxy of fatigue: as users progress through the study, they are more likely to be tired.
Adding \textit{`cumulative\_time\_spent'} as a feature causes large accuracy drops (22\%). The model is shifted toward \listening and \speaking, instead of \reading and \listening. The confusion matrix in Figure~\ref{fig:fatigue_cm} shows that \textit{`cumulative\_time\_spent'} causes the model to draw a large portion of the decisions on the \speaking and \listening. 

Both results for \textit{`engagement'} and \textit{`cumulative\_time\_spent'} are observed to have substantial influences on the model performances.
Although we took measures to avoid fatigue by maintaining the sessions short, this result suggests that shorter sessions may be more suitable (e.g., having a longer break between sections). However, further analysis using auxiliary signals  -- such as eye-blink frequency, average time to eye-closed duration, involuntary hand gestures -- will be used to validate whether \textit{`cumulative\_time\_spent'} can indeed infer fatigue.

\section{Conclusion}
\label{sec:discussion}


In this paper, we have used machine learning and the data collected in a pilot study ($N=7$) as a way to validate the robustness of our experimental design.
In particular, we conduct an ablation study to examine the influences of five variables by adding each variable as an additional feature along with the signal features. Based on the preliminary analysis from examining the changes in the model's classification performances, we can infer whether our experimental design needs further refinement before continuing with the study.
Given the exploratory nature of our investigation and small sample size limitations, further research is needed to fully understand the variables' impacts. 
But the methodology described in this paper allowed us to reveal the shortcomings in our experimental design early in data collection, and may assist practitioners in validating complex laboratory user study designs in a cost-effective manner.

\begin{acks}
This research is partially supported by 
the \grantsponsor{ARC}{Australian Research Council}{https://www.arc.gov.au/} (\grantnum{ARC}{CE200100005}, \grantnum{ARC}{DE200100064}).

\end{acks}

\newpage

\bibliographystyle{ACM-Reference-Format}
\balance
\bibliography{reference}

\end{document}

%% file: macros.tex
\newcommand{\ds}[1]{\textcolor{purple}{{\bf [Damiano: }{\em #1}{\bf ]}}}

\newcommand{\task}[1]{{\small\sffamily #1}\xspace}
\newcommand{\listening}{\task{LISTEN}}
\newcommand{\reading}{\task{READ}}
\newcommand{\writing}{\task{WRITE}}
\newcommand{\speaking}{\task{SPEAK}}
\newcommand{\relaxing}{\task{BASE}}